\documentclass[pra,showkeys,twocolumn]{revtex4}

\usepackage{amsmath}
\usepackage{amsfonts}            
\usepackage{amssymb}

\usepackage[colorlinks,urlcolor=blue,linkcolor=blue,pagecolor=blue,filecolor=blue,citecolor=blue]{hyperref}

\usepackage{mathrsfs}

\usepackage{eufrak}

\usepackage{verbatim} 

\usepackage{graphicx}
\usepackage{epstopdf}

\usepackage{bm}

\begin{document}


\title{Microsolvation of phthalocyanine molecules in superfluid helium nanodroplets as revealed by the optical line shape at the electronic origin}

\author{S. Fuchs}
\affiliation{Institut f\"{u}r Physikalische und Theoretische Chemie, Universit\"{a}t Regensburg, 93053 Regensburg, Germany}
\author{J. Fischer}
\affiliation{Institut f\"{u}r Physikalische und Theoretische Chemie, Universit\"{a}t Regensburg, 93053 Regensburg, Germany}
\author{M. Karra}
\affiliation{Fritz-Haber-Insitut der Max-Planck-Gesellschaft\\
Faradayweg 4-6, 14195 Berlin, Germany}
\author{B. Friedrich}
\email[Corresponding author: ]{bretislav.friedrich@fhi-berlin.mpg.de}%
\affiliation{Fritz-Haber-Insitut der Max-Planck-Gesellschaft\\
Faradayweg 4-6, 14195 Berlin, Germany}
\author{A. Slenczka}
\email[Corresponding author: ]{alkwin.slenczka@chemie.uni-regensburg.de}%
\affiliation{Institut f\"{u}r Physikalische und Theoretische Chemie, Universit\"{a}t Regensburg, 93053 Regensburg, Germany}

\date{\today}

\begin{abstract}
We investigate the solvent shift of phthalocyanine (Pc) doped into superfluid helium droplets  and probed by optical spectroscopy at the electronic origin. Our present work complements extant studies and provides results that in part contradict previous conclusions. In particular, the solvent shift does not increase monotonously with droplet radius all the way up to the bulk limit, but exhibits a turnaround instead. Moreover, a substructure is resolved, whose characteristics depend on the droplet size. This behavior can hardly be reconciled with that of a freely rotating Pc-helium complex.
\end{abstract}

\keywords{Helium droplets; electronic spectroscopy; phthalocyanine; microsolvation; line shape.}

\maketitle

\section{\label{sec:level1}Introduction\protect\\}

The optical line shape at the electronic origin of phthalocyanine doped into helium nanodroplets was the subject of our initial studies at the Regensburg helium droplet laboratory \cite{1, 2}. The remarkable asymmetry of the line shape, previously reported by Hartmann et al. \cite{3}, had been studied under variation of experimental parameters that determine the helium droplet size. By combining the excluded volume model \cite{4}, which accounts for shifts of electronic transition frequencies of a molecule surrounded by a polarisable environment, with the size distribution of singly-doped helium droplets, the asymmetric line shape could be simulated quantitatively. Moreover, this simulation could be inverted and used to infer the underlying droplet size distribution from a given asymmetry of the experimental line shape.

A frequency shift with respect to a gas-phase transition increases monotonously with increasing number of helium atoms involved and becomes observable already when a single helium atom is attached to a molecule. Given the finite effective reach of the London dispersion forces, the solvent shift is expected to converge to a maximum value corresponding to solvation in bulk helium. Under bulk conditions, the resonance frequency shift of an ensemble of solvated molecules is expected to be homogeneous. As reported by Lehnig et al. \cite{5}, bulk conditions for phthalocyanine are reached for droplets with an average number of 10$^6$ helium atoms or more. In such droplets the asymmetry of the line shape vanishes and a rather narrow double-peak feature appears.

Lehnig et al. \cite{5} were able to successfully fit this double-peak feature by the rotational band structure of an oblate symmetric top rotor with appropriately adapted moments of inertia: As a rule of thumb (RoTh), slow rotors, with a rotational constant of less than 1 cm$^{-1}$, exhibit moments of inertia in helium that are a threefold of those in the gas phase. This increase reflects a helium layer rigidly attached to the surface of the dopant molecule. Thereafter, a phthalocyanine solvation complex was identified by means of path integral Monte Carlo simulations (PIMC) \cite{6}. Alternative interpretations of the increased moments of inertia of Pc in helium by viscous drag of the rotating Pc molecule could be excluded by observing, via dispersed emission spectroscopy \cite{7, 8, 9, 10}, dynamical processes of the solvation layer induced by electronic excitation. Up to that point, experimental observations of Pc in helium droplets and the corresponding simulations were found to agree quantitatively with each another.

The effect of the droplet size distribution on the asymmetry of the line shape as well as the emergence of the rotational band structure once the inhomogeneous line broadening vanished for large droplets should be quite generally observable for any dopant species.

Our observation of the same double peak structure of phthalocyanine in large helium droplets ($>10^6$ atoms) as in Ref. \cite{5} but with a different intensity profile came as a surprise. A change in the intensity profile of the rotational band reflects either a change in the temperature or an altered rotor symmetry. According to the current understanding of helium droplets as cryogenic matrices for molecules with a fixed temperature of 0.37(1) K \cite{11}, the above possibilities are to be excluded. The recently recorded spectra, discussed below, of Pc in helium droplets comprised of 10$^4$-10$^7$ atoms are inconsistent with the assignment of the observed features as due to a rotational band structure. Hence free rotation of a phthalocyanine solvation complex in helium droplets is either unexpectedly different from that of other solvated molecules or does not occur.

\section{Experimental}

The experimental equipment consists of a vacuum machine with two differentially pumped vacuum chambers. The first chamber contains the helium droplet source which is a copy of the continuous flow nozzle developed in G\"{o}ttingen \cite{3}. It is equipped with a platinum orifice of 5 $\mu$m in diameter. The nozzle is attached to a Sumitomo cold head RDK-408S2 and compressor unit F-50Hw which provides cooling of the nozzle down to 7.0 K.

The second vacuum chamber contains the pick-up unit for doping of the helium droplets and the fluorescence detection unit and is accessed by the helium droplet beam via a conically shaped skimmer with a 1.4 mm diameter. Measured from the nozzle, the distance to the skimmer is about 20 mm and to the pick-up unit about 120 mm. The pick-up unit consists of a stainless steel cylinder surrounded by a heating wire. It is about 30 mm in diameter and 20 mm high. The heating wire is shielded by a closely contacted stainless steel cover and in addition by a copper tube connected to a liquid nitrogen Dewar. Additional 80 mm behind the pick-up unit a laser beam intersects the helium droplet beam at right angles. Orthogonal to both beam axes a condenser lens (f\# = 2) collects the laser induced fluorescence which is imaged onto the photocathode of a photo multiplier tube (PMT) (Hamamatsu R943-02), which is shielded by an appropriate edge filter in order to eliminate laser stray light. The PMT signal is amplified (two stages of SRS 445) and fed into a photon counter (SRS 400).

The laser system is an actively stabilized single mode ring dye laser (Coherent 899-29 autoscan) with a bandwidth of less than 1 MHz. It is pumped by 10 W @ 532 nm from an optically-pumped semiconductor laser (Coherent Verdi G10) and operated with DCM dye. The output power of the dye laser operated in single mode peaks at about 1 W.  In most of the experiments the laser is attenuated to only 1\% of the full power.

Frequency stepping of the laser and photon counting is synchronized by hardware hand-shaking between the laser and the photon counter. The management of the data reading and storing is accomplished by computer control via homemade software.

\section{Results and Discussion}

The asymmetric line shape at the electronic origin of phthalocyanine in helium droplets arises from the convolution of the log-normal droplet size and Poisson pick-up distributions with the droplet-size-dependent shifts of the transition frequencies involved, cf. Refs. \cite{1, 2}. The parameters of the log-normal droplet size distribution generated by the helium droplet source under subcritical expansion conditions have been determined empirically and listed in Ref. \cite{12}. The weighting of the \textit{nascent} droplet size distribution by the Poisson pick-up distribution \cite{13} for a single-molecule doping (the case considered here) gives rise to an \textit{effective} droplet size distribution.

\begin{figure}
\includegraphics[width=8.5cm]{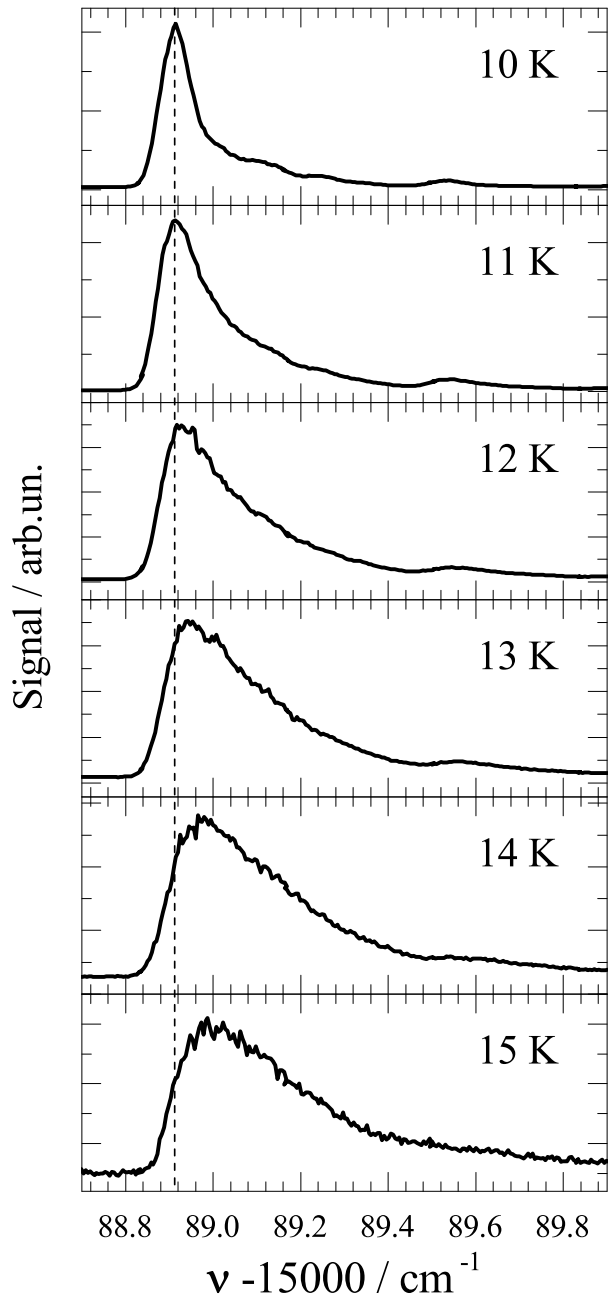}
\caption{\label{fig:1} Electronic band origin of phthalocyanine in superfluid helium droplets under variation of the temperature of the droplet source as indicated in each panel. From top to bottom the effective droplet size decreases. All spectra are recorded under subcritical He droplet source conditions.}
\end{figure}

In order to simulate the line shape function, the effective droplet size distribution has to be convoluted with the frequency shifts imparted by He droplets of a varying size. These individual-size shifts are due to the London dispersion forces between the dopant and the He droplet environment and reflect their long range character: the shifts depend on the droplet size only if the range of the London dispersion forces exceeds the size of the droplets. Hence with increasing droplet size, the frequency shift converges to the bulk value -- over 10$^6$ atoms for Pc. As a result, the effective droplet size distribution for singly-doped droplets leads to an asymmetric line shape with a tail extending towards the corresponding gas phase resonance frequency, which is blue-shifted for Pc (S$_0$-S$_1$), but can be either blue- or red-shifted for other transitions and dopants. Upon increasing the effective droplet size, the asymmetry vanishes and the line shape approaches that observed under bulk conditions. A quantitative model \cite{1,2} based on the above considerations has been shown to capture the behavior of phthalocyanine in helium droplets well. Figure \ref{fig:1} shows a series of electronic band origins of Pc recorded at a stagnation pressure of 20 bar and nozzle temperatures varying from 10 to 15 K, corresponding to a size variation of the droplets from $2\times10^4$ to $3.5\times10^3$ He atoms.

Of particular interest in the present study is the line shape recorded under quasi-bulk conditions \cite{5}, i.e., for droplets comprised of more than 10$^6$ helium atoms. Since the droplet size distribution generated by the helium droplet source -- whether sub- or supercritical -- spans a rather broad range, only a fraction of the distribution is capable of picking up single Pc molecules and thereby contribute significantly to the signal from singly-doped helium droplets. By tuning the molecular vapour pressure/density in the pick-up unit, the effective size distribution of singly-doped helium droplets can be shifted within the nascent droplet size distribution as generated by the droplet source. Alternatively, the effective droplet size distribution can be varied by tuning the temperature of the helium droplet source. A third option for varying the nascent distribution, namely by tuning the stagnation pressure in the helium source, has been omitted because of the concomitant rise of the background pressure in the vacuum chamber.

Therefore, in order to study the influence of the droplet size on the line shape, we varied either the partial pressure in the pick-up unit or the nozzle temperature of the helium droplet source, while keeping all other parameters constant.

\begin{figure*}
\includegraphics{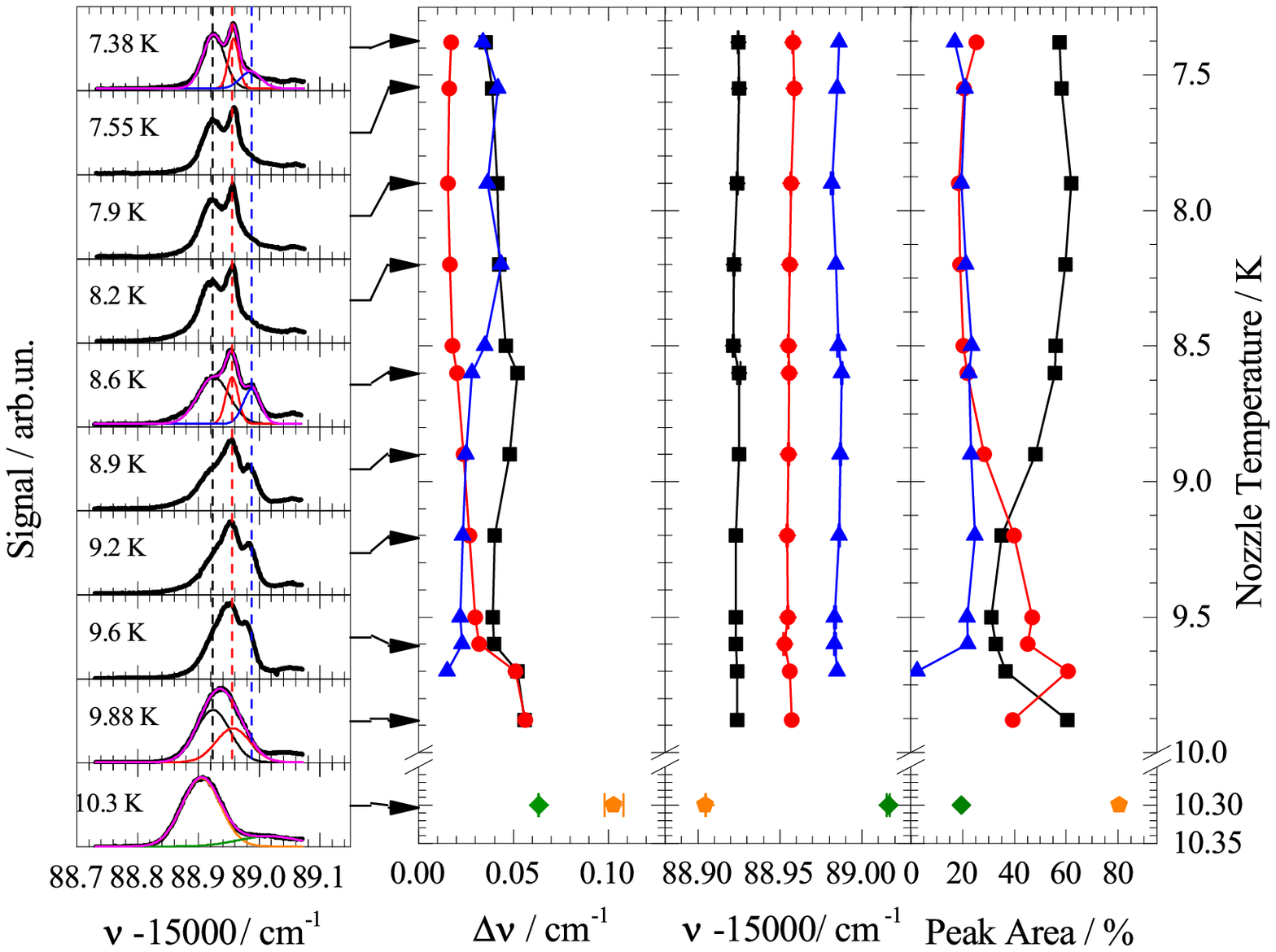}
\caption{\label{fig:2} High-resolution optical spectra at the electronic origin (S$_0$-S$_1$) of phthalocyanine in superfluid helium droplets taken at different values of the droplet source temperature, T, as indicated in each of the ten panels in the left column. The first panel on the right shows the half width (FWHM), $\Delta\nu$, the third panel the peak area of three Gaussian fits of the line shapes with almost constant peak positions, $\nu-15000$ (shown in the second panel), throughout the entire temperature range examined. The pick-up oven was heated by 8.5 W to a temperature of 540 K.}
\end{figure*}

The variation of the substructure resolved in the optical spectra of Pc at the electronic origin as observed at different droplet source temperatures is shown in Figure \ref{fig:2}. The droplet source temperature, T, ranges from 7.3 K to 10.2 K, while the stagnation pressure and the heating power of the pick-up unit were held constant at 20 bar and 8.5 W, respectively. Thus, the spectra shown pertain to varying effective droplet size, decreasing from top to bottom. At the lowest temperature of 7.3 K -- which corresponds to the largest droplets in this series, consisting of 10$^7$ He atoms -- a double peak feature is resolved (black and red peaks in the top panel of Fig. \ref{fig:2}) which closely resembles that reported in Ref. \cite{5}. Upon increasing the nozzle temperature, see the lower panels, the first peak on the red side (shown in black) decreases monotonously and becomes unresolved in the lowest spectrum. The minimum of its intensity, cf. the right-most panel of Fig. \ref{fig:2}, occurs at T=9.5 K where the expansion conditions change from supercritical (T$<9.5$ K) to subcritical (T$>9.5$ K). At the same time, a third, blue-shifted, peak (shown in blue) appears with increasing nozzle temperature and then disappears again as T$>9.6$ K. Finally, at a nozzle temperature of about 9.8 K, the substructure vanishes entirely and a single Gaussian-shaped peak remains whose peak position is shifted to the red with respect to the range of the triple-peak substructure, see the bottom-most panel. In addition, another peak (shown in green) could be fitted to the weak signal on the blue side which, for lower nozzle temperatures, also lies outside the frequency range of the triple peak. The high symmetry of the intense single peak observed at T=10.3 K (bottom panel) suggests that the corresponding He droplet is still rather large (with 10$^5$ atoms or more). As shown in Fig. \ref{fig:1}, the line shapes become asymmetric for smaller droplets.

The triple peak substructure, which is most clearly resolved for a nozzle temperature of 8.5 K, has not been reported before. The dependence of the relative intensities of the three peaks on the effective size distribution of the singly-doped droplets amounts to evidence against the interpretation of the rotational band structure as due to the free rotation of the phthalocyanine solvation complex:  For as long as the droplet size exceeds that of the dopant, the intensity profile of a rotational band should not depend on the droplet size.

\begin{figure*}
\includegraphics{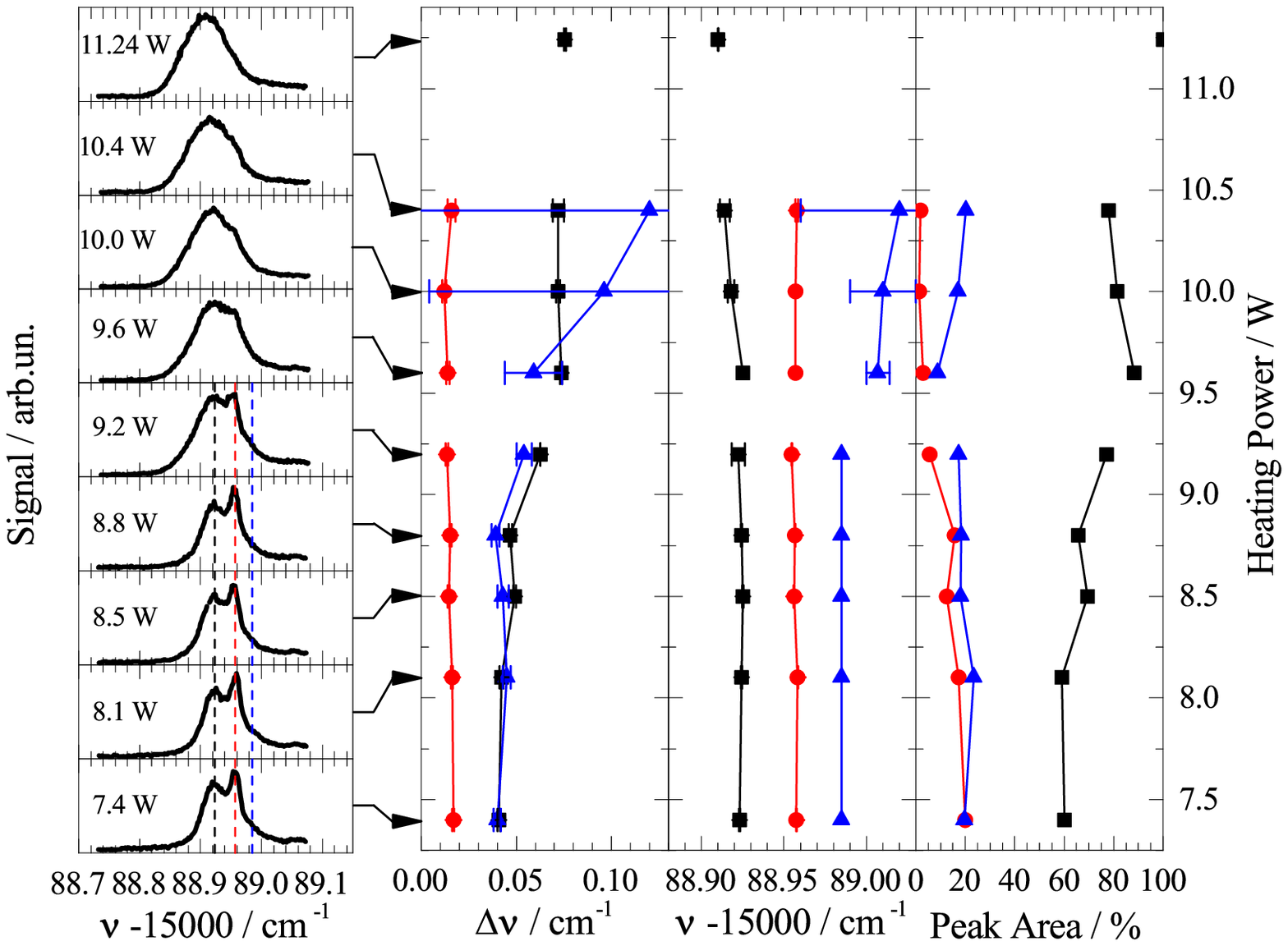}
\caption{\label{fig:3} High-resolution optical spectra at the electronic origin (S$_0$-S$_ 1$) of phthalocyanine in superfluid helium droplets taken at different vapour pressures of Pc in the pick unit as determined by its heating power. The droplet source temperature and pressure were held constant at the supercritical values of 8 K and 20 bar, respectively. Thus, from top to bottom, the effective droplet size distribution of singly-doped He droplets increases. The spectra were fitted with Gaussians whose half width (FWHM) ($\Delta\nu$), peak position ( $\nu$-15000), and intensity (peak area) are shown in the right three panels, respectively.}
\end{figure*}

In order to cross-check the results displayed in Fig. \ref{fig:2}, we measured the optical spectra at the band origin of Pc  in singly-doped He droplets at a fixed nascent droplet size distribution and varied the vapor pressure of Pc in the pick-up unit instead. The variation of the effective droplet size distribution is achieved via the droplet-size dependent single-molecule capture cross section. By choosing a nascent distribution centered at the droplet size of 10$^7$ He atoms (generated at supercritical values of source pressure, 20 bar, and temperature, 8 K \cite{11}), we ensured overlap with the range of conditions that led to some of the spectra shown in Fig. \ref{fig:2}.

The spectra obtained at a fixed source pressure and temperature but for varying heating power of the pick-up unit and thus Pc vapour pressure are displayed in Figure \ref{fig:3}. The heating power decreases from top to bottom, which means that the effective droplet size distribution for single-molecule doping shifts towards larger droplets within the given size distribution generated by the droplet source. The top spectrum in Fig. \ref{fig:3} still shows a singly-peaked band that could be fitted by a single Gaussian. The symmetry of the peak is indicative of the proximity to the bulk limit, i.e., the droplet size is at least 10$^5$ helium atoms. Clearly, this single Gaussian peak resembles closely that shown at the bottom of Fig. \ref{fig:2} despite of the different nascent droplet size distribution.

By reducing the heating power of the pick-up cell and thus the vapour pressure of Pc, a single Pc molecule will be captured by ever larger droplets. Thus the series of spectra shown in Fig. \ref{fig:3} corresponds to increasing effective droplet size from top to bottom. With increasing effective droplet size a substructure appears in the spectra that could be fitted, like the substructures in Fig. \ref{fig:2}, by a trio of Gaussians, whose half widths (FWHM) ($\Delta\nu$), peak positions ($\nu$), and intensities (peak areas) are shown in Fig. \ref{fig:3} in the panes to the right of the spectra. The upper four spectra differ from the lower five: while the peak positions of the lower five spectra are identical to those of the trio of Gaussians identified in Fig. \ref{fig:2}, the upper four spectra have different peak positions which are spread over a larger spectral range; moreover, the intensity of the middle peak becomes negligible. We note that in order to mask the influence on the fitting process of the signal that is spreading further to the blue, the third peak on the blue side of the five lower spectra had to be kept fixed at its initial position of 15088.985 cm$^{-1}$.  In the case of the upper four spectra, this restriction was relaxed. Thus, the third peak on the blue side shifts towards the signal which is outside the intense leading structure. In the top spectrum, the main part of the signal shifts towards the red and a second peak (not shown in Fig. \ref{fig:3}) coincides with that part of the signal which for the rest of the spectra was outside the spectral range of the fitting procedure. This is similar to what was shown in the bottom spectrum in Fig. \ref{fig:2}, cf. the green peak. Throughout the lower five spectra in Fig. \ref{fig:3}, the intensity profile is dominated by the first two peaks of the trio while the third one appears as a small bump at the blue side. Nevertheless, its intensity, as given by the peak area, is not negligible, cf. the right-most pane in Fig. \ref{fig:3}. The first two peaks (black and red) correspond to the double-peak feature assigned in Ref. \cite{5} to the rotational band structure of the solvation complex of phthalocyanine in a helium droplet. However, as shown by the series of spectra displayed in Fig. \ref{fig:3}, the intensity profile varies with the effective droplet size distribution. This behavior, again, thwarts the assignment of the observed substructure to the rotational fine structure.

This invalidation of the assignment of the observed substructure to a rotational band system is of consequence for the understanding of microsolvation of phthalocyanine in helium droplets. This goes hand in hand with the observed solvent shift which, instead of converging to a maximum value, undergoes a turnaround upon crossing a certain droplet size, cf. the two bottom spectra in Fig. \ref{fig:2} and the first top spectra in Fig. \ref{fig:3}. Both of these observations reveal a behavior that is inconsistent with our understanding of free rotation of a dopant-helium solvation complex in a superfluid helium droplet.

The explanation of the turnaround of the solvent shift can be attempted via two alternative model considerations: (A) Within the excluded volume model, the turnaround could be justified by a reduction of the effective radius of the polarizable helium environment surrounding the dopant species, despite the increasing size of the polarizable environment (droplet). As noted in Ref. \cite{14}, this could be accomplished by a reduction of the helium density, conceivable in the close vicinity of the solvation complex. Such a density reduction can be expected for a solvation complex captured by a vortex inside a superfluid helium droplet. However, the appearance of vortices requires the excess of a droplet size of 10$^8$ atoms \cite{15}, which is not available under our experimental conditions. (B) Alternatively, one can consider the state-specific solvation energy. For the observed red shift, the solvation energy is larger for the excited state than for the ground state. In addition, either of these energies increases monotonously with the radius of the polarizable environment and finally converges to the bulk limit. In case this limit is reached at different radii, namely, a smaller of the two for the excited state, the resonance energy would undergo a turnaround from a red shift to a blue shift. However, it appears rather unlikely that a stronger solvent shift would converge at a smaller radius than a weaker one.

\begin{figure}
\includegraphics[width=8.5cm]{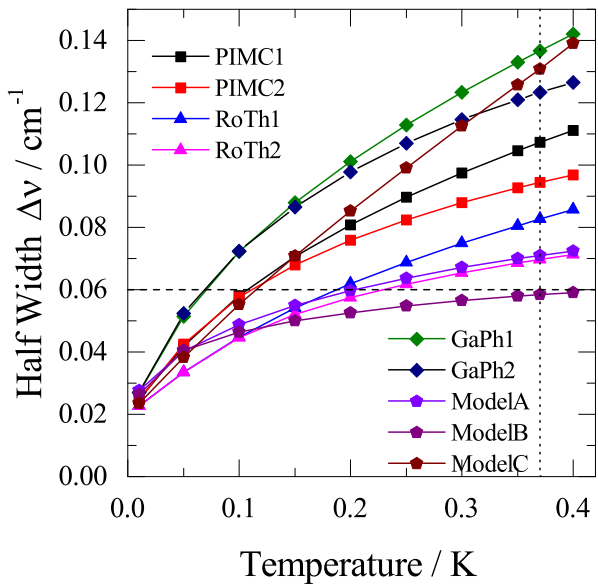}
\caption{\label{fig:4} Half width (FWHM) of the rotational fine structure as a function of temperature obtained from simulations for nine rotor models (cf. Table \ref{tab:table1}). The helium droplet temperature of 0.37 K is marked by a vertical dotted line. PIMC and RoTh make use of rotational constants obtained from path integral Monte Carlo simulations [6] and from Ref. [5], respectively; GaPh pertains to bare phthalocyanine. Indices 1 and 2 distinguish between electronic transitions without and with the change of the moments of inertia, respectively. Finally, ModelA and ModelC simulate the change from symmetric top to spherical top and ModelB a gradual change from symmetric towards spherical top. The horizontal dashed line marks the upper limit of the half widths (FWHM) shown in Figs. 2 and 3, first pane to the right of the spectra.}
\end{figure}

Taking the rotational fine structure reported in Ref. \cite{5} as a starting point, one may wonder about what it would take to assign a rotational fine structure - apparently a different one - to each of the various substructures observed in the experiment. As these experimentally observed substructures differ in their peak positions, peak areas, and even the number of peaks when examined in He droplets of varying size, a rational assignment of the rotational fine structure should not exceed the width of any one of the three Gaussian peaks.

The parameters to vary in order to fit the rotational band structure to the experimental results consist of two sets of rotational constants (pertaining to the ground -- double prime -- and excited -- single prime -- states) and the temperature of the system. We tested two models, one based on the structure of a phthalocyanine-helium solvation complex as identified by path integral Monte Carlo (PIMC) simulations \cite{6} and another on the empirical structure used in Ref. \cite{5} (RoTh). In addition, we also considered bare Pc \cite{22} (GaPh) as a third model, for the sake of comparison. For each of the three models the calculations have been carried out twice - with and without a rather marginal change of the rotational constants upon electronic excitation of Pc (indicated by 1 and 2 respectively). In addition, three more cases (ModelA/B/C) have been tested, starting with the symmetric rotor as obtained from PIMC simulations and ending with a perfect (A and C) or half way (B) spherical top rotor. The values of the rotational constants used are listed in Table \ref{tab:table1} and the resulting half widths (FWHM) of the rotational band systems are plotted as a function of the rotational temperature in Figure \ref{fig:4}. For the droplet temperature of 0.37 K all models except of ModelB exhibit half widths larger than the upper limit obtained in the experiment marked by the dashed horizontal line in Fig. \ref{fig:4}. Even though the half width of ModelB is less than the Gaussian fits to the experimental data, one should keep in mind that these Gaussian fits do not correspond to rotational bands. Thus, the half width of the rotational band has to be significantly smaller than the half width of any of the Gaussians. Hence, one can see that it is only the temperature effect that is capable of squeezing the width of the rotational fine structure into a range which might be hidden below the half widths shown in Figs. \ref{fig:2} and \ref{fig:3}. We note that, so far, no other temperature of helium droplets than 0.37 K, marked by a vertical dotted line in Fig. \ref{fig:4}, has been found. This temperature is determined by the process of evaporative cooling whose time constant of about 1 ns is short compared to any time scales of the experiment.

\begin{table}
\caption{\label{tab:table1} Rotational constants in units of 10$^{-3}$ cm$^{-1}$ as used in calculations of the half widths of the rotational band system, see text and Fig. \ref{fig:4}.}
\begin{ruledtabular}
\begin{tabular}{ccccccc}
System &A"&B"&C"&A'&B'&C'\\
\hline
PIMC1& 1.72	& 1.72	& 1.02	& 1.72	& 1.72	& 1.02 \\
PIMC2& 1.72	& 1.72	& 1.02	& 1.65	& 1.65	& 1.06\\
RoTh1& 1.0	& 1.0	& 0.5	& 1.0	& 1.0	& 0.5\\
RoTh2& 1.0	& 1.0	& 0.5	& 0.96	& 0.96	& 0.52\\
ModelA& 1.72	& 1.72	& 1.02	& 1.72	& 1.72	& 1.72\\
ModelB& 1.72	& 1.72	& 1.02	& 1.72	& 1.72	& 1.37\\
ModelC& 1.72	& 1.72	& 1.02	& 1.02	& 1.02	& 1.02\\
GaPh1& 2.98	& 2.97	& 1.49	& 2.98	& 2.97	& 1.49\\
GaPh2& 2.98	& 2.97	& 1.49	& 2.86	& 2.85	& 1.55\\
\end{tabular}
\end{ruledtabular}
\end{table}

\section{Conclusion}

In light of the convincing account of microsolvation of phthalocyanine in helium droplets based on previous investigations of the line shape of the corresponding electronic origin \cite{1, 2, 5}, our present results come as a surprise. In particular, our detailed study of the electronic origin under variation of the effective droplet size distribution reveals two effects that elude explanation within the framework of the previous work: (1) A turnaround of the solvent shift was found which is not expected in the excluded volume model \cite{4}; the solvent shift of the electronic transition energy of a molecule is expected to increase monotonously with the radius of a polarizable environment and to converge to the bulk limit. (2) For rather large droplets generated under supercritical expansion conditions, a substructure was clearly resolved whose intensity profile could be fitted by three slightly overlapping Gaussian peaks. Moreover, the half width and the peak area of the three Gaussians were found to change upon variation of the effective droplet size. This behavior of the trio of Gaussians is inconsistent with the assignment to a rotational fine structure of the phthalocyanine solvation complex in a superfluid helium droplet. Under the given experimental conditions and for a size-independent droplet temperature such a rotational fine structure should exhibit an intensity profile independent of the effective droplet size distribution.

Apparently, the process of free rotation of a solvation complex inside a superfluid helium droplet is not always as simple as revealed by rotationally-resolved IR spectra. Since  modifications of the rotor parameters over a wide range have failed to squeeze the rotational band structure of a phthalocyanine solvation complex within the observed Gaussian peaks, this may indicate that there is no free rotation upon electronic excitation of phthalocyanine in helium droplets after all.

Within the context of the present study on phthalocyanine in helium droplets, kindred work on tetracene \cite{14} and on glyoxal \cite{16} is of particular interest. In the case of tetracene in helium droplets, a double peak was resolved at the electronic origin which could not be explained as due to a rotational band structure. Moreover, the second of these peaks on the blue side exhibits a rich fine structure consisting of numerous sharp peaks that are inconsistent with a plausible rotor model of the corresponding solvation complex either. To date, in helium droplets only glyoxal is known to exhibit a fully resolved rotational fine structure at the electronic origin. Analysis of the spectra revealed unexpectedly large changes in the moments of inertia upon electronic excitation of the solvated molecule, of up to +61\%. As stated in Ref. \cite{16}, this is due to a change of the electron density distribution. Moreover, glyoxal in helium droplets is the only species studied so far whose phonon wing coincides with the spectrum of elementary excitations of superfluid helium \cite{20}. The exceptional behavior of glyoxal may be indicative of a rather marginal perturbation of the rotating solvation complex and of its coupling to the helium droplet body upon electronic excitation. In contrast, for tetracene, phthalocyanine, and most other dopant species in helium droplets (cf. for example Ref. \cite{17, 18, 21}), the rotational band system is absent and the observed phonon wing appears  unrelated to the spectrum of elementary excitations of superfluid helium. Such behavior points to much stronger perturbations due to the change of the electron density distribution that accompanies electronic excitation. As noted earlier \cite{19}, electronic spectroscopy of molecules in helium droplets is particularly sensitive to changes of the electron density distribution, very little of which can be seen in the corresponding spectra of bare molecules.

More experimental data are needed to clarify this question. Our study will continue with complementary investigations of other molecules such as porphyrin. In addition, polar derivatives such as AlCl-phthalocyanine will be investigated by means of fluorescence excitation spectroscopy in a Stark field with the goal of further elucidating the effects of the superfluid He droplet environment on the rotation of such large dopant species.

We gratefully acknowledge financial support by the Deutsche Forschungsgemeinschaft (DFG) through the SPP1807 program.


\end{document}